\documentclass[pra,aps,superscriptaddress,showpacs,reprint,longbibliography]{revtex4-1}

\usepackage{etoolbox}
\usepackage{gensymb}
\usepackage{hyperref}
\usepackage{amsmath}
\usepackage{amssymb}
\usepackage{dsfont}
\usepackage[T1]{fontenc}
\usepackage{ae}
\usepackage{color}
\usepackage{graphicx} 
\usepackage{units}

\begin{document}

\title{Measuring the dispersive frequency shift of a rectangular microwave cavity \newline 
	induced by an ensemble of Rydberg atoms}

\author{M.~Stammeier} \email{mathiass@phys.ethz.ch} \affiliation{Department of Physics, ETH Z\"urich, CH-8093 Z\"urich, Switzerland}

\author{S.~Garcia} \affiliation{Department of Physics, ETH Z\"urich, CH-8093 Z\"urich, Switzerland}

\author{T.~Thiele} \affiliation{Department of Physics, ETH Z\"urich, CH-8093 Z\"urich, Switzerland}

\author{J.~Deiglmayr} \affiliation{Laboratorium f\"ur Physikalische Chemie, ETH Z\"urich, CH-8093, Z\"urich, Switzerland}

\author{J.~A.~Agner} \affiliation{Laboratorium f\"ur Physikalische Chemie, ETH Z\"urich, CH-8093, Z\"urich, Switzerland}

\author{H.~Schmutz} \affiliation{Laboratorium f\"ur Physikalische Chemie, ETH Z\"urich, CH-8093, Z\"urich, Switzerland}

\author{F.~Merkt} \affiliation{Laboratorium f\"ur Physikalische Chemie, ETH Z\"urich, CH-8093, Z\"urich, Switzerland}

\author{A.~Wallraff} \affiliation{Department of Physics, ETH Z\"urich, CH-8093 Z\"urich, Switzerland}

\renewcommand{\i}{{\mathrm i}} \def\1{\mathchoice{\rm 1\mskip-4.2mu l}{\rm 1\mskip-4.2mu l}{\rm
1\mskip-4.6mu l}{\rm 1\mskip-5.2mu l}} \newcommand{\ket}[1]{|#1\rangle} \newcommand{\bra}[1]{\langle
#1|} \newcommand{\braket}[2]{\langle #1|#2\rangle} \newcommand{\ketbra}[2]{|#1\rangle\langle#2|}
\newcommand{\opelem}[3]{\langle #1|#2|#3\rangle} \newcommand{\projection}[1]{|#1\rangle\langle#1|}
\newcommand{\scalar}[1]{\langle #1|#1\rangle} \newcommand{\op}[1]{\hat{#1}}
\newcommand{\vect}[1]{\boldsymbol{#1}} \newcommand{\id}{\text{id}}

\raggedbottom

\begin{abstract}
In recent years the interest in studying interactions of Rydberg atoms or ensembles thereof with optical and microwave frequency fields has steadily increased, both in the context of basic research and for potential applications in quantum information processing. We present measurements of the dispersive interaction between an ensemble of helium atoms in the $37\text{s}$ Rydberg state and a single resonator mode by extracting the amplitude and phase change of a weak microwave probe tone transmitted through the cavity. The results are in quantitative agreement with predictions made on the basis of the dispersive Tavis-Cummings Hamiltonian. We study this system with the goal of realizing a hybrid between superconducting circuits and Rydberg atoms. We measure maximal collective coupling strengths of $1~\text{MHz}$, corresponding to $3\cdot 10^3$ Rydberg atoms coupled to the cavity. As expected, the dispersive shift is found to be inversely proportional to the atom-cavity detuning and proportional to the number of Rydberg atoms. This possibility of measuring the number of Rydberg atoms in a nondestructive manner is relevant for quantitatively evaluating scattering cross sections in experiments with Rydberg atoms.  
\end{abstract} 	
\maketitle

\section{Introduction}
\label{sec:intro}

Light-matter interaction enhancement in cavity quantum electrodynamics (QED) leads to many interesting phenomena. 
The first demonstrations of this enhancement with Rydberg atoms in the microwave domain~\cite{Goy1983} and with ground-state atoms in the optical domain~\cite{Raizen1989} were followed by demonstrations in solid-state systems including superconducting qubits~\cite{Wallraff2004}, quantum dots~\cite{Reithmaier2004}, nitrogen vacancy centers~\cite{Park2006} and magnons in a YIG sphere~\cite{Tabuchi2014a}. Rydberg atoms constitute a particularly interesting type of emitter, owing to their their long lifetimes and large transition dipole moments 
for transitions in the microwave domain. Moreover, the combination of optical and microwave transitions could provide a path to realize quantum frequency conversion between the optical and the microwave domain~\cite{Andrews2014}. 

Cavity QED systems contribute significantly to the rapidly expanding field of quantum information processing~\cite{Imamog1999, Zheng2000, Kimble2008}. 
In cavity QED, quantum nondemolition (QND) measurements provide ideal projective measurements of either part of the system, emitter or photon, by using the other to acquire information~\cite{Grangier1998, Guerlin2007, Lupascu2007}. These measurement schemes enabled many experiments at the core of quantum physics, such as the observation of the quantum Zeno effect~\cite{Bernu2008,Barontini2015} or the implementation of quantum feedback~\cite{Smith2002,Sayrin2011,Vijay2012}. They are further used to generate squeezed states of the many-atom pseudo-spin~\cite{Schleier2010,Cox2016} that allow metrological measurements below the standard quantum limit.
For approximate two-level systems, QND measurements are commonly performed in the dispersive regime, where the interaction leads to a mutual dispersive shift in energy for the cavity and the emitter. 
In circuit QED, for instance, measurements of the cavity dispersive shift are the standard tool for qubit-state detection, achieving single-shot readout within hundreds of nanoseconds~\cite{Jeffrey2014}.  
In Rydberg cavity QED experiments, measuring the atomic dispersive shift allows one to prepare and detect the quantum state of the cavity field non-destructively~\cite{Nogues1999, Varcoe2000}.
Using the cavity dispersive shift to determine the state of a single Rydberg atom or a Rydberg-atom ensemble, however, is less explored. 
Maioli et al.~\cite{Maioli2005} have measured the phase shift of a coherent cavity field resulting from the dispersive interaction with a few Rydberg atoms, which allowed them to non-destructively determine either the atom state or the atom number. This measurement was performed with a coherent tone that mapped the phase shift to an intensity difference, which was then, in a second step, measured with a mesoscopic number of resonant atoms. 
While, in the optical domain, cavity transmission measurements were employed to observe the vacuum Rabi mode splitting and the dispersive shift caused by a beam of atoms~\cite{Raizen1989} or single atoms~\cite{Mabuchi1999} traversing the cavity, to our knowledge no such measurements have been reported for Rydberg atoms in the microwave domain.

In this article, we present measurements of the dispersive shift of a 3D cavity induced by an ensemble of Rydberg atoms. In Section~\ref{sec:experimental}, we introduce the experimental setup, which allows for the preparation and detection of helium  Rydberg atoms and for the detection of a cavity shift by transmission measurements with low probe-photon number. 
In Section~\ref{sec:results}, we present our results on the time-dependent dispersive shift and the dispersively shifted cavity spectrum. We show that the system is quantitatively described by the dispersive Tavis-Cummings Hamiltonian for an ensemble of two-level atoms and a single cavity mode. 
Moreover, we study the dispersive shift under controlled variation of the system parameters: the atom number and the atom-cavity detuning. Finally, in Section \ref{sec:discussion}, we discuss the potential of the experiment for precise, nondestructive atom-number and qubit-state measurements.   

\section{Experimental setup}
\label{sec:experimental}

\subsection{Rydberg-atom preparation and detection}

\begin{figure}[b] \centering \includegraphics[width=80mm]{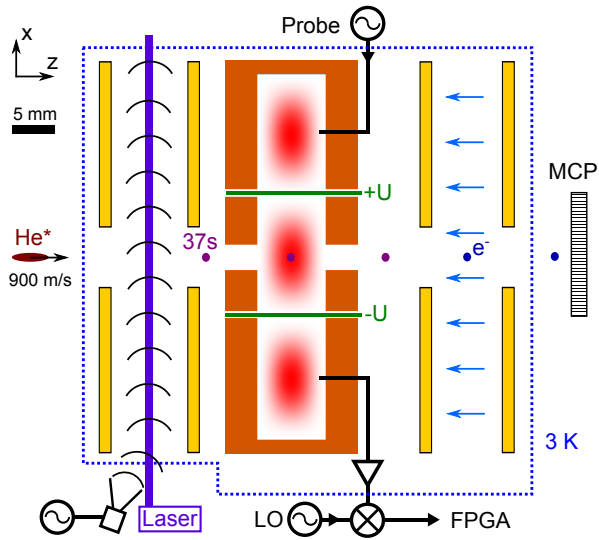} 	
\caption{ Schematic diagram of the experimental setup: a rectangular 3D cavity (orange) is mounted between two pairs of circular electrodes (yellow). The $\text{TE}_{301}$ mode field intensity is depicted in red and the embedded electrodes are drawn in green. Blue arrows represent the pulse field used for field ionization of the Rydberg atoms. The dimensions inside the cryogenic environment (blue dots) are roughly to scale.  }
\label{fig:Setup} \end{figure}

In the setup (see~\cite{Thiele2014} for details) sketched in Fig.~\ref{fig:Setup}, in each experimental cycle, we prepare Rydberg atoms, have them interact with the cavity and detect them. 
Within a high-vacuum chamber, a liquid-nitrogen-cooled pulsed valve with incorporated dielectric barrier discharge~\cite{Even2015} generates a cloud of supersonic ${{}^{4}\text{He}}^{*}$ atoms in the metastable singlet state ($1\text{s}^{1}2\text{s}^{1}~{}^{1}S_{0}$). The atom cloud, traveling at a speed of $v_{z}=900 \pm 13 ~\text{m}\cdot{\text{s}}^{-1}$, then enters an ultrahigh vacuum, cryogenic environment at $3~\text{K}$, where a rectangular copper 3D cavity is mounted between two pairs of circular electrodes. Between the first pair of electrodes, a small fraction of the ${\text{He}}^{*}$ atoms is photoexcited to the $37\text{p}$ Rydberg state ($1\text{s}^{1}37\text{p}^{1}~{}^{1}P_{1}$ with lifetime $\tau_{37\text{p}} \simeq 2.7~ \mu \text{s}$~\cite{Theodosiou1984}) by a $10$-$\text{ns}$-long UV laser pulse with a wavelength of $313~\text{nm}$. The Rydberg atoms are then coherently transferred to the longer-lived $37\text{s}$ state (lifetime $\tau_{37\text{s}} \simeq 45~ \mu \text{s}$) using a $250$-$\text{ns}$-long microwave pulse with frequency close to the field-free transition frequency $\omega_{\text{a,0}} / 2\pi= 21.5299~\text{GHz}$.

The resulting ensemble of about 4000 $37\text{s}$ Rydberg atoms then propagates through the cavity (described in Appendix~\ref{app:cavity}), where it interacts with the central maximum of the critically coupled $\text{TE}_{301}$ mode with resonance frequency $\omega_{\text{c}} / 2\pi = 21.532~\text{GHz}$ and total cavity decay rate $\kappa / 2\pi = 4.1~\text{MHz}$. 
Inside the cavity, two electrodes installed parallel to the atomic beam allow us to tune the atomic transition frequency $\omega_{\text{a}}$ via the quadratic dc Stark effect by applying electric potentials of opposite polarity ($+U$ and $-U$ with respect to the grounded cavity).

After leaving the cavity, the Rydberg atoms are ionized with a pulsed field at the second pair of electrodes and the resulting electrons are detected on a microchannel plate (MCP). The signal from the MCP, current amplified and integrated on a digital oscilloscope, constitutes a relative measure of the number of Rydberg atoms.

\subsection{Detection of cavity frequency shifts}
\label{subsec:method}

\begin{figure}[b] \centering \includegraphics[width=80mm]{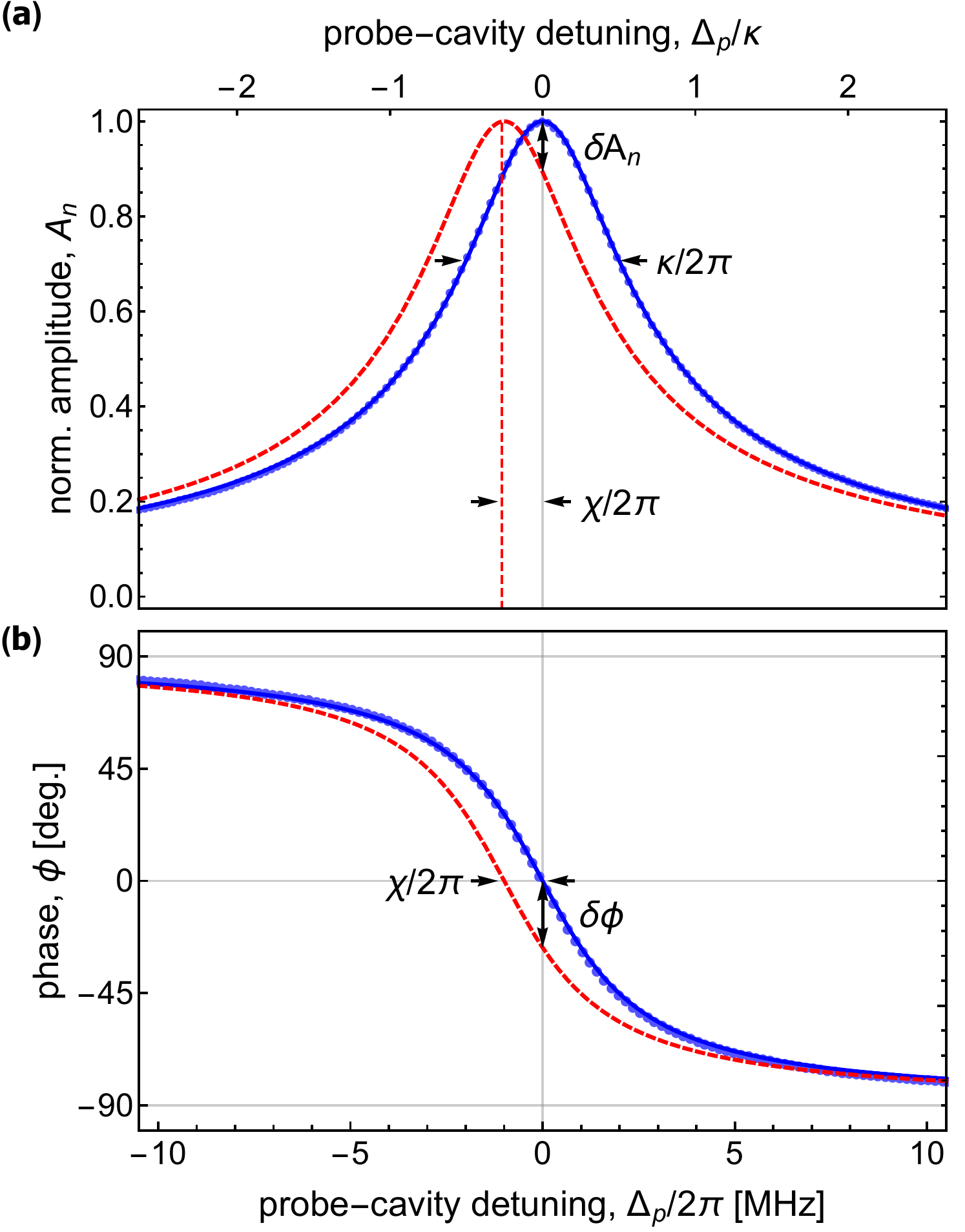} 	
	\caption{(a) Normalized amplitude $A_{\text{n}}$ and (b) phase $\phi$ of the measured transmission spectrum. The transmission of the $\text{TE}_{301}$ mode was measured (blue points) and fitted (blue solid line). Dashed red lines show transmission spectra with frequency shift of $\chi / 2\pi=-1~\text{MHz}$, exaggerated for clarity.} 
	\label{fig:spectrum}
\end{figure}	

We characterize the resonator transmission spectrum of the $\text{TE}_{301}$ mode in amplitude (normalized to its maximum) and phase with a vector network analyzer (see Fig.~\ref{fig:spectrum}).
The data shows the expected Lorentzian resonator transmission. The dependence on the probe frequency $\omega_{\text{p}}$, or the probe-cavity detuning $\Delta_{\text{p}}=\omega_{\text{p}}-\omega_{\text{c}}$, is fitted with the normalized amplitude $A_{\text{n}}(\Delta_{\text{p}}) =1/ \sqrt{1+{\left(2\Delta_{\text{p}} / \kappa \right)}^{2}}$
 and phase $\phi(\Delta_{\text{p}}) =-\arctan \left(2\Delta_{\text{p}} / \kappa \right)$ to obtain the values for $\omega_{\text{c}}$ and $\kappa$ stated above.
We detect the shift in resonance frequency of the cavity induced by the atom ensemble as a change in transmission amplitude $\delta \! A_{\text{n}}$ and phase $\delta \phi$. The observable amplitude and phase changes are dependent on the probe-cavity detuning as indicated in Fig.~\ref{fig:spectrum} with a cavity frequency shift exaggerated by a factor of about 10 for clarity.

To observe the dispersive frequency shift induced by the Rydberg atoms passing through the cavity, we continuously measure the cavity transmission. For this purpose, we apply a weak probe tone that induces a coherent state of about $600$ photons in the cavity, as calculated from the probe power and input-output theory~\cite{Gardiner1985}. For the detection, we amplify the transmitted signal with a cryogenic low-noise amplifier and employ heterodyne down-conversion by mixing with a local oscillator. The complex envelope of the signal $\mathcal{A}(t)$ is recorded following digital homodyne down-conversion and filtering on a field-programmable gate array (FPGA). The detection has a time resolution of approximately $100~\text{ns}$ dominated by the digital filtering. The noise background is characterized by $34$ effective noise photons referred to the cavity output \cite{Eichler2013a}.  When the atoms have left the cavity, we take a reference trace $\mathcal{A}_0(t)$ that allows us to calculate the change in amplitude $\delta \! A_{\text{n}}(t)$ and phase $\delta \phi(t)$, which are further averaged to reduce the noise, with typically $5 \cdot 10^4$ repetitions of the experiment at $25~\text{Hz}$ repetition rate (Appendix~\ref{app:noiseave} provides details about noise and averaging).

\section{Measurements of cavity dispersive shift}
\label{sec:results}

\subsection{Time-dependent dispersive shift 
                  and dispersively shifted transmission spectrum}
\label{ssec:results2}

\begin{figure*}[t] \centering \includegraphics[width=160mm]{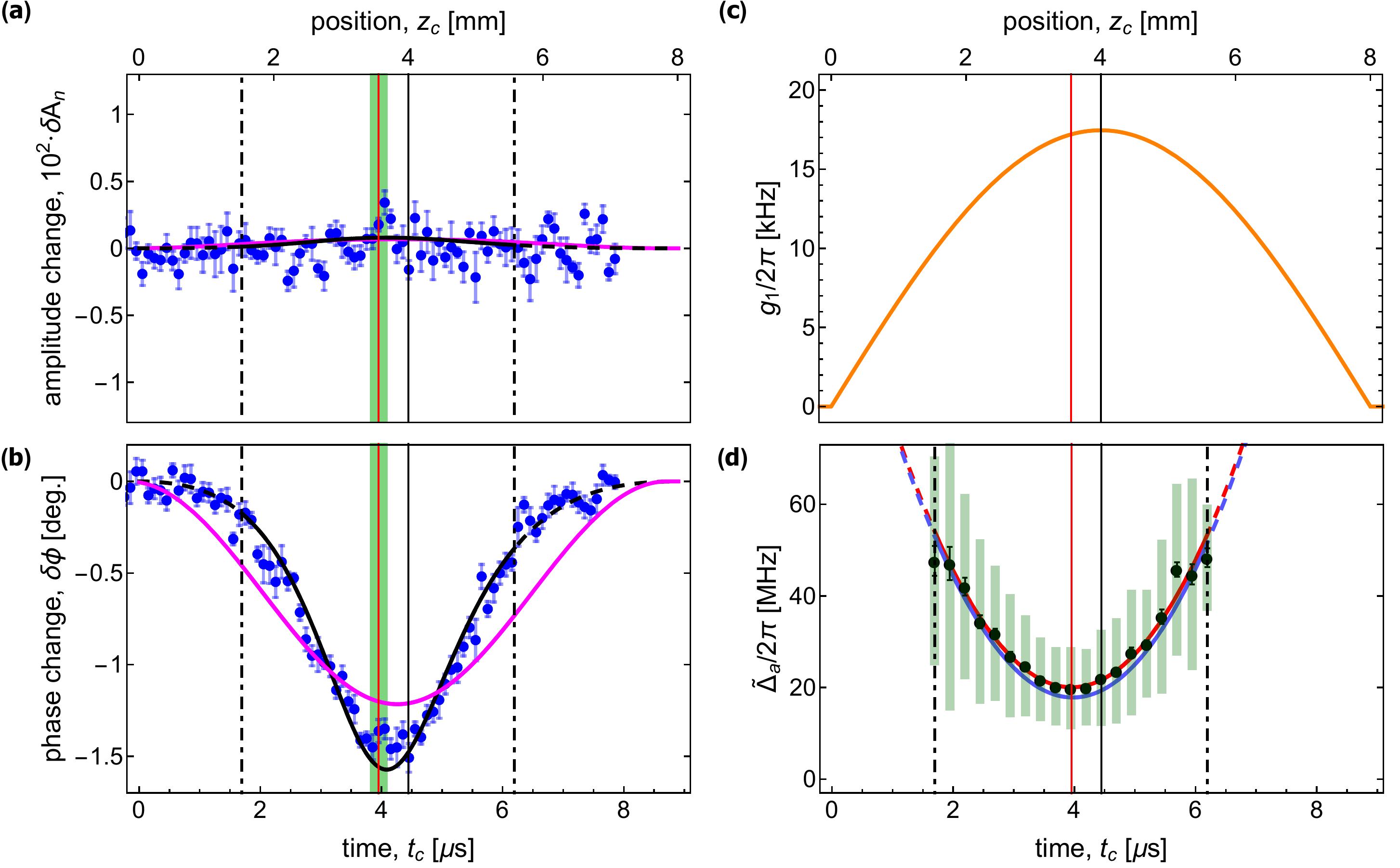} 	
\caption{Change in (a) normalized amplitude $\delta \! A_{\text{n}}$  and (b) phase $\delta \phi$ as a function of position $z_{\text{c}}=v_{\text{z}}\cdot t_{\text{c}}$ (along the cavity axis, given by the time the atoms travel inside the cavity $t_{\text{c}}$) measured with zero probe-cavity detuning $\Delta_{\text{p}}$ (blue points). The full lines represent fits of the data assuming a constant atom-cavity detuning (purple) and based on the measured time dependence of the atom-cavity detuning (black).
The cavity center is marked by a black vertical line in all plots and the green bars indicate a $300~\text{ns}$ time window used for averaging. 
 (c) Calculated single-atom coupling strength $g_{\text{1}}$ (orange) as a function of $z_{\text{c}}$.
(d) Measurement of the raw atom-cavity detuning $\tilde{\Delta}_{\text{a}}$ (black points) as a function of $t_{\text{c}}$. The red parabola is a fit of the measured raw atom-cavity detunings and the red vertical line indicates the position of the minimal atom-cavity detuning. The blue parabola displays the atom-cavity detuning $\Delta_{\text{a}}$ that is used for the fit in (a) and (b)~(see text for details). 
Dashed dotted lines indicate the measurement range. Black error bars represent the standard error on the center frequency and green bars the full widths at half maximum of the fitted atomic spectral lines. 
}
 \label{fig:modelplot}
 \end{figure*}

\begin{figure}[t] \centering \includegraphics[width=80mm]{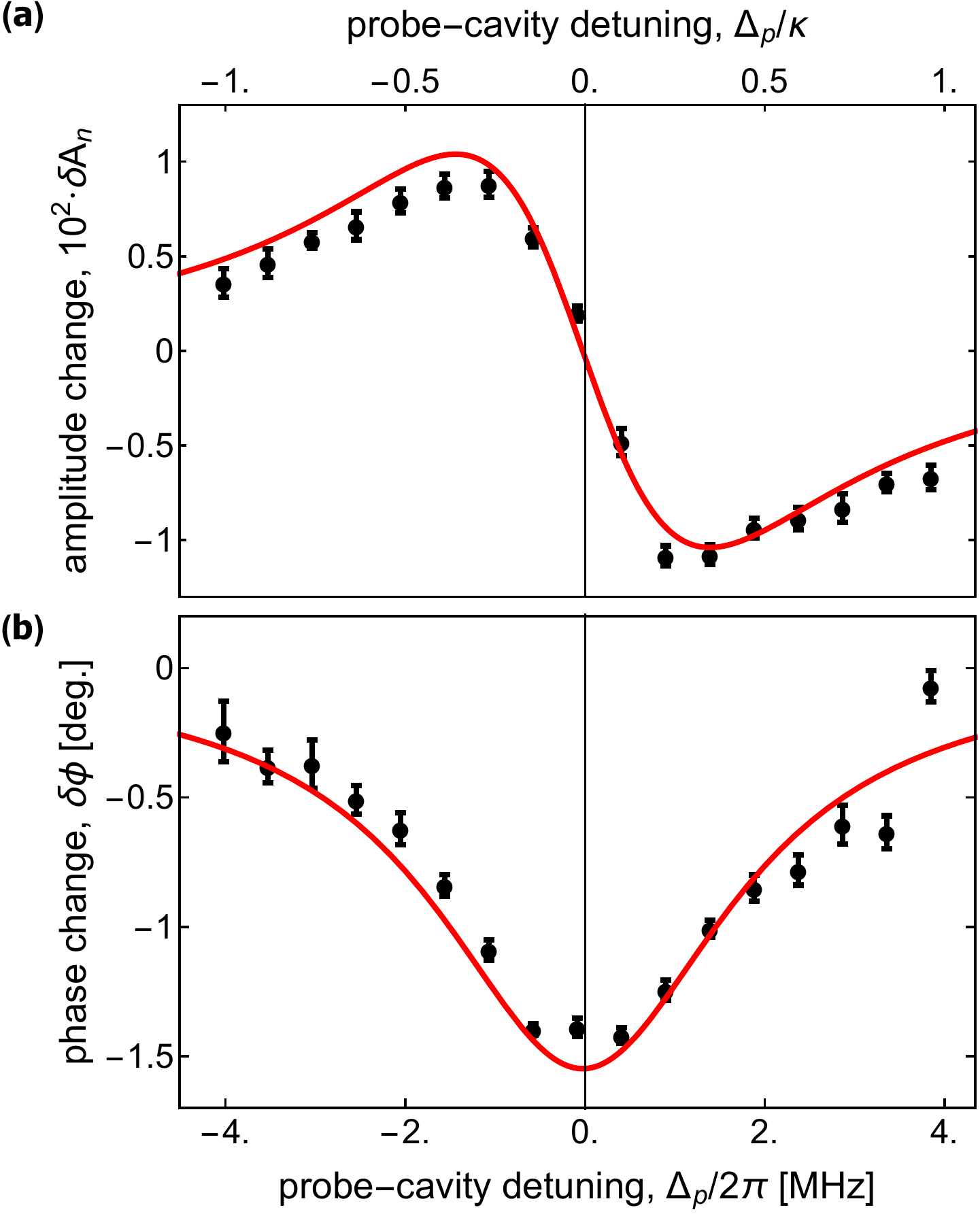} 	
\caption{ Change in (a) normalized amplitude $\delta \! A_{\text{n}}$  and (b) phase $\delta \phi$ as a function of the probe-cavity detuning $\Delta_{\text{p}}$ (black points). The error bars indicate one standard deviation. A fit of the data with the complete model is shown in red (see text).}
 \label{fig:freqscanplot}
 \end{figure}

We have measured the time-dependent change of cavity transmission for probe-cavity detunings in the frequency range $\left(-\kappa,\kappa \right)$ with $\kappa / 8$ steps (full data set presented in Appendix~\ref{app:fulldata}). In Fig.~\ref{fig:modelplot}(a) and (b), we show the normalized amplitude change $\delta \! A_{\text{n}}(t_{\text{c}})$ and the phase change $\delta \phi(t_{\text{c}})$ at the cavity resonance frequency ($\Delta_{\mathrm{p}} = 0$), as functions of the position                
$z_{\text{c}}=v_{z}\cdot t_{\text{c}}$ along the cavity axis, given by the time $t_{\text{c}}$ elapsed since the Rydberg atoms have entered the cavity. We observe little amplitude change $\delta \! A_{\text{n}}(t_{\text{c}})$. In contrast, the phase change $\delta \phi(t_{\text{c}})$ gradually decreases from zero at the cavity entrance to its minimal value of ${\delta \phi}_{\text{min}}=-1.39(6)~\text{deg.}$ at the cavity center (black line at $z_{\text{m}}=4~\text{mm}$) and then increases back to zero as the atoms move towards the end of the cavity. A symmetric time dependence of the phase change is expected from the spatial microwave field distribution of the $\text{TE}_{301}$ mode. From ${\delta \phi}_{\text{min}}$ and the cavity decay rate $\kappa$, we extract a first estimate of the maximal dispersive shift ${\langle \chi \rangle}_{\text{max}} / 2\pi=-49(2)~\text{kHz}$. 

To fully analyze the experimental results, 
we model the Rydberg-atom cloud as a pointlike ensemble of $N$ two-level atoms with identical $37\text{s}-37\text{p}$ transition frequency $\omega_{\text{a}}$ and identical single-atom coupling $g_{\text{1}}$ to the microwave field.  
The Tavis-Cummings Hamiltonian describes the system~\cite{Tavis1968}. In the dispersive limit, where the atom-cavity detuning $\Delta_{a}=\omega_{a}-\omega _{c}$ is much larger than the collective coupling strength $g_{N}=g_{\text{1}}\cdot \sqrt{N}$, this leads to a collective cavity dispersive shift~\cite{Blais2007,Zueco2009}
\begin{equation}
\chi = \frac{{g_{N}}^2}{\Delta_{\text{a}}} \cdot \frac{J_{z}}{N/2} \text{,} \label{dispshift1}
\end{equation}
which depends on the pseudo-spin polarization $J_{z}$ of the ensemble.
The latter is given as the sum of the individual pseudo-spin operators, $\sigma_{z}^{i}/2$, of the $37\text{s}-37\text{p}$ transition, $J_{z}=\frac{1}{2}\sum_{i=1}^{N}\sigma_{z}^{i}$, 
and has an expectation value  $\langle J_{z} \rangle = -N / 2$, when all atoms have been prepared in the spin-down state $37\text{s}$. This leads to a collective dispersive shift 
 \begin{equation}
\langle  \chi \rangle= -\frac{{g_{1}}^2 N}{\Delta_{\text{a}}}.   \label{dispshift2}
\end{equation} 
A model for the change in amplitude $\delta \! A_{\text{n}}$ and phase $\delta \phi$ is then obtained from the difference between the dispersively shifted and the unperturbed transmission spectrum. The model depends on constant parameters such as the probe-cavity detuning $\Delta_{\text{p}}$ and the number of Rydberg atoms $N$, but also on the single-atom coupling strength $g_{1}$ and the atom-cavity detuning $\Delta_{\text{a}}$, which in turn depend on the position of the atoms in the cavity and are thus time-dependent.

In a first step, we assume a constant atom-cavity detuning and use the time dependence of the calculated single-atom coupling shown in Fig.~\ref{fig:modelplot}(c). The latter emerges from the variation of the mode function over the length $l$ of the cavity in the beam direction and the atomic velocity $v_{z}$. The coupling is $g_{1}(t_{\text{c}}) / 2\pi= 17.5~\text{kHz} \cdot \sin\left(\pi v_{z}t_{\text{c}} / l \right)$, as determined from the $37\text{s}-37\text{p}$ transition dipole moment ($1092~e a_{0}$, calculated according to~\cite{Zimmerman1979}) and the analytic $\text{TE}_{301}$ mode function. Fitting this simple model to the data (magenta lines in Fig.~\ref{fig:modelplot}(a) and (b)) provides a correct description of the data, but also reveals the necessity of taking into account the time dependence of the atom-cavity detuning for a quantitative description.

We have therefore independently measured the atom-cavity detuning within the cavity along the propagation axis by coherently transferring the population from the $37\text{s}$ to the short-lived $37\text{p}$ state with a $200~\text{ns}$ microwave pulse~(see~\cite{Thiele2014} for details). The coherent transfer (at time $t_{\text{c}}$) results in fewer Rydberg atoms being detected at the MCP, and thus in a spectral line, when the frequency of the microwave pulse is scanned over the atomic transition frequency. A Gaussian fit of the spectral line determines the raw atom-cavity detuning  $\tilde{\Delta}_{\text{a}}=\Delta_{\text{a}}+\delta_{\text{ac}}(z_{\text{c}})$, which differs from the real atom-cavity detuning $\Delta_{\text{a}}$ by a small, position-dependent ac Stark shift $\delta_{\text{ac}}(z_{\text{c}})$ (induced by the strength of the microwave pulse ~\cite{Bohlouli-Zanjani2007}).
The fit further determines the full width at half maximum of the atomic spectrum, which includes the contribution from the Fourier-limited width ($5~\text{MHz}$) of the $200$-ns-long microwave pulses.
Using this method, we have mapped out the raw atom-cavity detuning $\tilde{\Delta}_{\text{a}}$ as a function of $t_{\text{c}}$ (see Fig.~\ref{fig:modelplot}(d)) between  $1.7~\mu\text{s}$ and $6.2~\mu\text{s}$ (black dash-dotted lines) in steps of $250~\text{ns}$.
The time dependence of the raw atom-cavity detuning is fitted with a parabola $\tilde{\Delta}_{\text{a}}(t_{\text{c}}) / 2\pi \simeq 6.65~\frac{{\text{MHz}}^2}{{\mu \text{s}}^2} \cdot {(t_{\text{c}}-t_{\text{min}})}^{2}+20.07~\text{MHz}$ (red), the minimum ($t_{\text{min}} \simeq 3.96~\mu \text{s}$, red line) of which is shifted by approximately $450~\mu \text{m}$ from the cavity center. While the parabolic dependence can readily be explained by linearly increasing stray electric fields towards the cavity walls, the spatial shift points to higher stray fields at the cavity exit, where impurities and charges in the atomic beam are more likely to collect on the cavity wall. 

Subtracting the position-dependent ac Stark shift $\delta_{\text{ac}}(z_{\text{c}})=\delta_{\text{ac}}(z_{\text{m}})\cdot {\sin}^{2}\left(\pi z_{\text{c}} / l \right)$, where $\delta_{\text{ac}}(z_{\text{m}}) / 2\pi=2.4(4)~\text{MHz}$ is independently measured at the cavity center, we obtain a model for $\delta \! A_{\text{n}}(t_{\text{c}})$ and $\delta \phi(t_{\text{c}})$ that takes into account the time dependences of both the atom-cavity detuning $\Delta_{\text{a}}(t_{\text{c}})$ and the collective atom coupling $g_{N}(t_{\text{c}})$. Fitting this model to the data (black lines in Fig.~\ref{fig:modelplot}(a) and (b)) yields excellent quantitative agreement over the full data set (see Appendix~\ref{app:fulldata}). The fit determines the two free parameters of the model: the origin of the time scale ($t_{\text{c}}=0~\mu s$), which is used to align the data with the cavity frame, and the number of coupled Rydberg atoms $N =3.3(2) \cdot 10^3 $, which corresponds to a maximal collective coupling of $g_{N\text{,max}} / 2\pi=1.01(2) ~\text{MHz}$. 

The difference between the cavity spectra with and without atoms is depicted in Fig.~\ref{fig:freqscanplot}. Here, the data for each probe-cavity detuning corresponds to the mean of $\delta \! A_{\text{n}}(t_{\text{c}})$ and $\delta \phi(t_{\text{c}})$ in the cavity center (averaging over $300~\text{ns}$ around $t_{\text{min}}$, green bar in Fig.~\ref{fig:modelplot}(a) and (b)). 
The data (black) is consistent with the probe-cavity detuning dependences $\delta \! A_{\text{n}}(\Delta_{\text{p}})$ and $\delta \phi(\Delta_{\text{p}})$ of the  complete model (red). The small dispersive shift ${\langle \chi \rangle}_{\text{max}} / \kappa\simeq 1\%$ implies that $\delta \! A_{\text{n}}(\Delta_{\text{p}})$ and $\delta \phi(\Delta_{\text{p}})$ are proportional to the negative derivatives of the unperturbed amplitude and phase response $A_{\text{n}}(\Delta_{\text{p}})$ and $\phi(\Delta_{\text{p}})$. We have therefore clearly observed a collective dispersive shift, the time dependence of which is determined by the time-dependent collective coupling $g_{N}(t_{\text{c}})$ and atom-cavity detuning $\Delta_{\text{a}}(t_{\text{c}})$. 

The observed dispersive shift $\chi$ depends on the probe power: it tends to vanish when the number of photons $n_{\mathrm{c}}$ inside the cavity increases, due to higher-order terms that are neglected in the dispersive approximation. 
For a single two-level emitter (qubit), the probe-power dependence of the dispersive shift is negligible for photon numbers much lower than the critical photon number $n_{\mathrm{crit}} = \Delta_{\mathrm{a}}^2 / 4 g_1^2$~\cite{Blais2004}. When $n_{\mathrm{c}}$ reaches the order of $n_{\mathrm{crit}}$, the observed dispersive shift decreases significantly and limits the signal-to-noise ratio of the qubit-state read-out ~\cite{Boissonneault2008}. This effect has been investigated in circuit QED experiments~\cite{Gambetta2006}. In Appendix~\ref{app:ncrit}, we show that the critical photon number for the Tavis-Cummings Hamiltonian of $N$ qubits is identical with $n_{\mathrm{crit}}$ of the single-qubit case, if $N \ll n_{\mathrm{crit}}$ holds.  In our system, $n_{\mathrm{crit}} \simeq 10^5 $ and $ N \simeq 10^3$, so $n_{\mathrm{crit}}$ still constitutes the threshold around which the dispersive shift starts to reduce. However, to obtain a precise estimate of the atom number from our measurements, we choose a probe-photon number ($n_{\mathrm{c}}\simeq 600$) two orders of magnitude lower than $n_{\mathrm{crit}}$. In Appendix~\ref{app:ncrit}, we present measurements of the probe-photon number dependence  and show that the dispersive shift remains unaffected under these conditions.

\subsection{Variation of system parameters}
\label{ssec:results3}

The dispersive shift depends on the collective coupling strength $g_{\text{N}} $ and the atom-cavity detuning $\Delta_{\text{a}}$, which further characterize the dispersive Tavis-Cummings Hamiltonian. We independently control both parameters in our experiment. We exploit the linearity of $\delta \phi(\Delta_{\text{p}}=0)$ with the dispersive shift for $\chi \ll \kappa$ and report here mainly measurements of phase change at zero probe-cavity detuning, where the data is averaged in the $300~\text{ns}$ time window around the time of minimal atom-cavity detuning $t_{\text{min}}$. 

\begin{figure}[t] \centering \includegraphics[width=76mm]{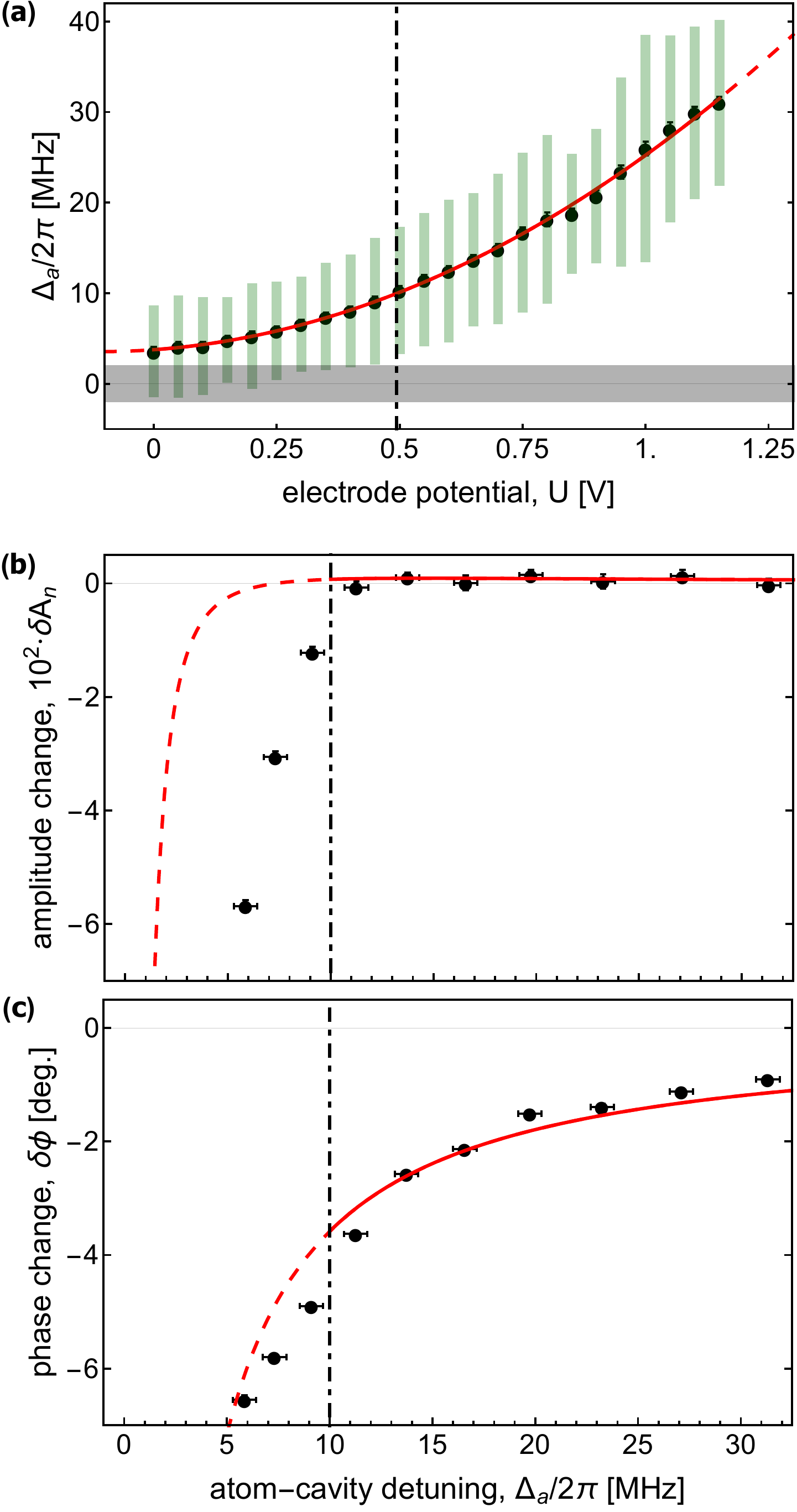} 	
	\caption{ (a) Ac Stark shift corrected measurement of the atom-cavity detuning $\Delta_{\text{a}} / 2\pi$ (black points) at $t=t_{\text{min}}$ as a function of the potential $U$ applied to the cavity electrodes and fit of the quadratic Stark effect with a parabola (red). The gray and green bars represent the full widths at half maximum of the cavity and the atomic spectral lines, respectively.  
		Change in (b) normalized amplitude $\delta \! A_{\text{n}}$  and (c) phase $\delta \phi$ at zero probe-cavity detuning as a function of the measured atom-cavity detuning $\Delta_{\text{a}} / 2\pi$ shown in (a). The error bars indicate one standard deviation for data averaged over $300~\text{ns}$. The fit of the complete model (red) is limited to atom-cavity detunings above a threshold $\Delta_{\text{a,crit}} / 2\pi = 10~\text{MHz}$ (dash-dotted black line in all plots).}
	\label{fig:detuningscan}
\end{figure}

We first vary the atom-cavity detuning by applying potentials of opposite polarity ($+U$ and $-U$) to the two electrodes mounted within the cavity to induce a quadratic dc Stark shift, as depicted in Fig.~\ref{fig:detuningscan}(a), where the constant ac Stark shift $\delta_{\text{ac}}(z_{\text{m}})$ is already subtracted. The measured Stark shifts agree well with the fitted parabola displayed in red, which thus accurately describes the atom-cavity detuning $\Delta_{\text{a}}(U)$.
 The measured change in amplitude $\delta \! A_{\text{n}}$ and phase $\delta \phi$ is displayed as a function of the atom-cavity detuning in Fig.~\ref{fig:detuningscan}(b) and (c). While $\delta \! A_{\text{n}}$ stays approximately constant down to a critical detuning $\Delta_{\text{a,crit}} / 2\pi= 10~\text{MHz}$ (dash-dotted vertical lines in Fig.~\ref{fig:detuningscan}),  $\delta \phi$ follows the expected $1 / \Delta_{\text{a}}$ dependence. The fit (red), limited to atom-cavity detunings above $\Delta_{\text{a,crit}}$, shows reasonable agreement with the data for $N = 4.3(2) \cdot 10^3 $ Rydberg atoms. Below the critical detuning, the measured atomic spectral lines overlap with the cavity spectrum (compare the gray bar and the green bars in Fig.~\ref{fig:detuningscan}(a)) and a fraction of the atoms is resonantly excited to the $37\text{p}$ state, which decays rapidly, mainly to the ground state. This additional loss mechanism for the photons in the cavity explains the drop of $\delta \! A_{\text{n}}$ at atom-cavity detunings below $\Delta_{\text{a,crit}}$ (see Fig.~\ref{fig:detuningscan}(b)).  

\begin{figure}[t] \centering \includegraphics[width=80mm]{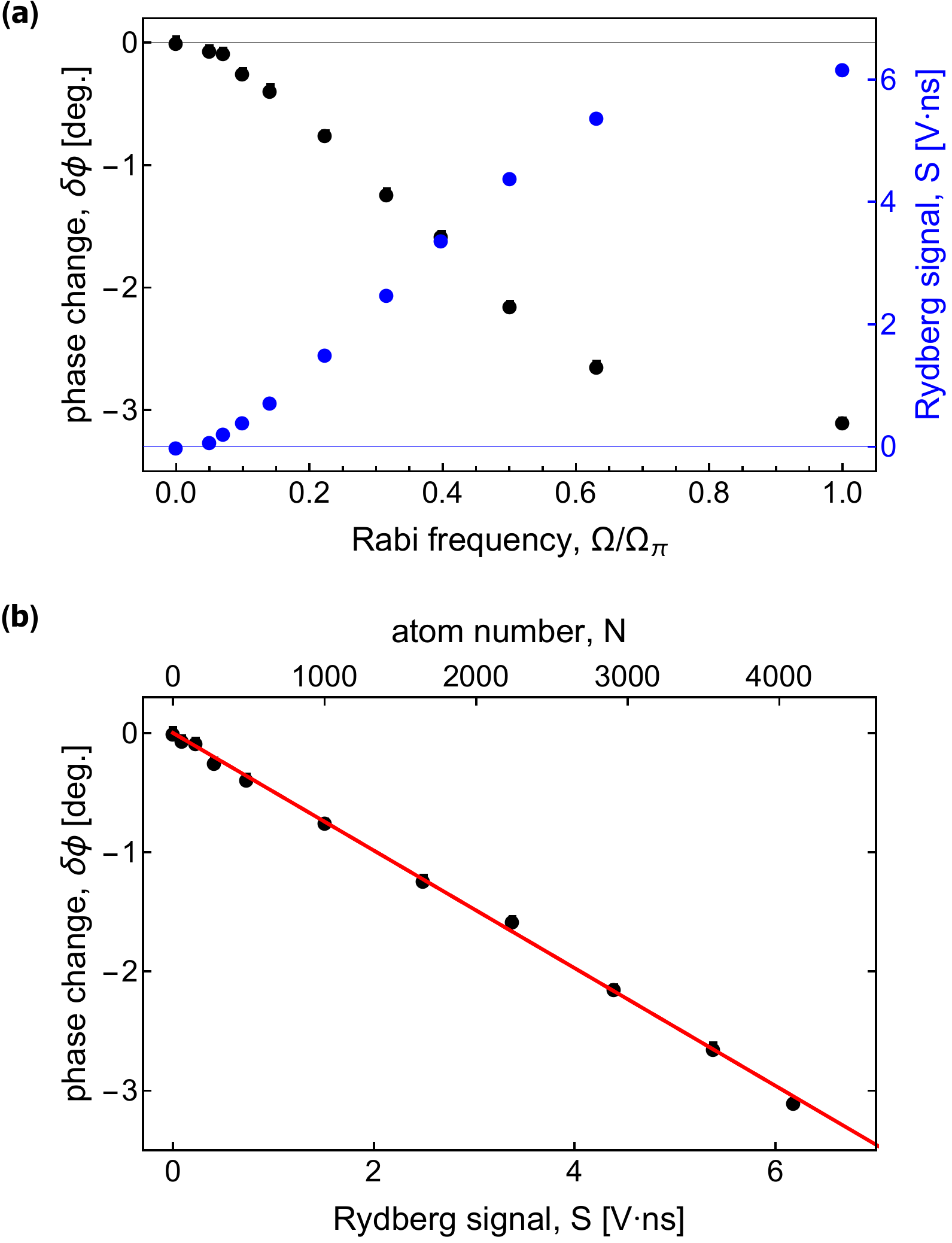} 	
	\caption{(a) Phase change $\delta \phi$ at zero probe-cavity detuning $\Delta_{\text{p}}$ (black vertical axis and points) and Rydberg signal $S$ (blue vertical axis and points) as a function of the normalized microwave pulse amplitude $\Omega / \Omega_{\pi}$. 
		(b) Phase change $\delta \phi$ versus Rydberg signal $S$. The fit of the complete model (red) to the data calibrates the atom number scale on the top axis.}
	\label{fig:atomscan}
\end{figure}

We then vary the number of Rydberg atoms $N$, using the amplitude of the preparation microwave pulse that coherently transfers the atoms from the $37\text{p}$ to the longer-lived $37\text{s}$ state. The integrated signal $S^*$ measured at the MCP is proportional to the total number of Rydberg atoms and has a constant contribution $S_0 = 0.79(3)~\text{V}\! \cdot \! \text{ns}$ from Rydberg states other than $37\text{s}$ (measured in the absence of the microwave pulse) and $37\text{p}$ (which decays by $\simeq 99\%$ on the way to the cavity center). Subtracting this contribution, we obtain the Rydberg signal $S=S^*-S_0$ that is proportional to the number of atoms in the $37\text{s}$ state. In Fig.~\ref{fig:atomscan}(a) we display $1 / 2$ of a complete Rabi cycle of $S$ (blue points and axis) as a function of the microwave pulse amplitude expressed as the Rabi frequency $\Omega$ normalized to the $\pi$-pulse frequency $\Omega_{\pi}$.  
In the same figure, we plot the phase change $\delta \phi$ (black points and axis) at the constant atom-cavity detuning $\Delta_{\text{a}} / 2\pi = 11.25(55)~\text{MHz}$ as a function of the normalized Rabi frequency $\Omega / \Omega_{\pi}$. We note that $\delta \phi$ is zero for $\Omega / \Omega_{\pi}=0$ and gradually decreases with increasing Rabi frequency and increasing number of atoms in the $37\text{s}$ state. We expect a linear dependence of $\delta \phi$ on $S$, because $\delta \phi$ (measured at zero probe-cavity detuning) is proportional to the dispersive shift, which in turn is expected to be proportional to the number of $37\text{s}$-state atoms. This linear dependence between dispersive shift and Rydberg-atom number resulting from the collective coupling is confirmed in Fig.~\ref{fig:atomscan}(b), where we plot $\delta \phi$ directly as a function of $S$. Here, the fit (displayed in red) estimates a maximal number of Rydberg atoms $N_{\text{max}} = 4.1(2) \cdot 10^3 $ for the Rydberg signal at $\Omega / \Omega_{\pi}=1$, which is then used to calibrate the atom number axis on top of the plot. In the presented results, the maximal number of $37\text{s}$ Rydberg atoms varies between $3.3(2) \cdot 10^3$ and $4.3(2) \cdot 10^3$, as a consequence of experimental conditions, i.e., mainly the efficiency of the source of metastable singlet atoms, which varied between the presented measurements (on the time scale of a week).

\section{Conclusions and Outlook}
\label{sec:discussion}

We have presented measurements of the microwave cavity dispersive shift induced by Rydberg atoms obtained from the cavity transmission with photon numbers well below its critical value. In particular, we have modeled the experimentally observed time dependence of the dispersive shift resulting from the interplay between the time-dependent collective coupling strength and atom-cavity detuning. We have also verified the dispersive shift by measurements of the cavity transmission as a function of the probe-cavity detuning. Controlling the parameters of the atomic cloud, we have shown that the collective dispersive shift $\chi$ is proportional to the number of Rydberg atoms $N$ and inversely proportional to the atom-cavity detuning $\Delta_{\text{a}}$. The results agree well with the dispersive Tavis-Cummings Hamiltonian, and consistently imply maximal collective coupling strengths above $g_{N\text{,max}} / 2\pi\simeq 1~\text{MHz}$, corresponding to more than $3300$ coupled Rydberg atoms.

With the method presented here, we are able to determine the atom number in a nondemolition measurement. We obtain a relative uncertainty in the atom number of $\sigma_N / N \sim 4\%$. However, this atom number is an effective quantity, because we neglect the variation of the single-atom coupling strength and the atom-cavity detuning across the dimension of the atom cloud by assuming a point-like ensemble. Indeed, in our experiment, the atom cloud is large enough (diameter $\simeq1~\text{mm}$) to be sensitive to the inhomogeneities of the microwave and static electric fields. Since these fields vary over the cavity length of $8~\text{mm}$, 
we expect a small deviation of the determined effective atom number from the real atom number. Assuming that the field inhomogeneities over the atom cloud can be suppressed in future experiments by using larger cavities, smaller atomic ensembles and transitions that are less sensitive to electric fields, with otherwise identical  parameters, the relative uncertainty in the atom number could reach $\simeq 0.3\%$.  

The ratio of dispersive shift and cavity decay rate $x = 2\chi / \kappa$ can be optimized for the best atom number determination in phase measurements on resonance: for small values of $x$, the uncertainty is large because the observed phase shift decreases as $\delta \phi = \mathrm{arctan} (x)$. For large values of $x$, the uncertainty is also substantial because the phase shift saturates and the transmitted signal amplitude decreases. The optimum ratio is estimated to be $x_{\mathrm{opt}} \simeq 0.8 $ (details in Appendix~\ref{app:SNR}), which leads to a relative precision in the determination of the atom number $\sigma_N / N \simeq 3.33/\sqrt{\mathrm{SNR} \, k}$. Here,  $k$ is the number of averages and the single-shot power $\mathrm{SNR} = n_{\mathrm{c}} \kappa_{\mathrm{out}} \tau / n_{\mathrm{noise}} $ is determined by the number of photons in the cavity $n_{\mathrm{c}}$, the coupling rate $\kappa_{\mathrm{out}}$ of the cavity to its output port, the integration time $\tau$ and the effective noise photon number $n_{\mathrm{noise}}$ of the detection. For a given cavity and atomic transition, the optimal ratio can be achieved by adjusting the atom-cavity detuning to 
$\Delta_{\mathrm{a,opt}} = 2g_N^2 / \left( x_{\mathrm{opt}}\kappa \right)$. 
In the dispersive approximation ($\Delta_{\mathrm{a}} \gg g_N \, , \, \kappa$), the optimal detection requires collective strong coupling $g_N \gg \kappa$, which can be obtained with a superconducting cavity, in which cavity losses are reduced by several orders of magnitude.  

The resolution might be further enhanced by improvements of the SNR, e.g., by using quantum-limited amplifiers ($n_{\mathrm{noise}}\simeq 1$), and slowing down the atoms to maximize $\tau$, for example with Rydberg-Stark deceleration~\cite{Vliegen2007,Allmendinger2013}. 
Using realistic parameters ($N=4000$, $g_{N} / 2\pi = 1.1 ~\text{MHz}$, $\kappa / 2\pi \simeq \kappa_{\mathrm{out}} / 2\pi = 300~\text{kHz}$ for an over-coupled cavity, $\Delta_{\mathrm{a,opt}} / 2\pi \simeq 10 ~\text{MHz}$ , $n_{\text{c}} = 880 \simeq 10^{-2} n_{\mathrm{crit}} $ and $\tau = 50~\mu \text{s}$), the single-shot power SNR increases by more than 3 orders of magnitude to $8 \cdot 10^4$. The relative and absolute uncertainties of the atom number then become $\sigma_N / N \simeq 1.2\%$ and $\sigma_N \simeq 49$  \textit{in a single-shot measurement}, which is below the width of a poissonian distribution ($\sqrt{N} \simeq 63$). Such precise nondemolition measurements of large atom numbers open up the prospect of determining absolute scattering and reactive cross sections in experiments with Rydberg atoms and molecules~\cite{Wrede2005,Dai2005,Allmendinger2016}. 

Another important potential application of the presented measurement method is the nondemolition detection of the quantum state of the atomic ensemble, characterized by its pseudo-spin $J_z$. This measurement requires a long lifetime of the excited state of the atomic transition. In helium atoms this could be achieved with s-p transitions ($\tau_{\text{p,n=50}} \sim 100 \  \mu \mathrm{s}$) of triplet Rydberg states or by using transitions involving high-angular-momentum Rydberg states ($\tau_{\text{l=49,n=50}} \sim 30 \ \mathrm{ms}$). 

Long qubit lifetimes are also relevant in the context of quantum memories, where Rydberg atoms could for instance act as a memory of the quantum state of a superconducting qubit in a hybrid cavity QED scheme~\cite{Petrosyan2009}. Coherent state transfer between both systems via virtual photons could be achieved within $1 \ \mu$s with the collective coupling measured here ($1~\text{MHz}$) and typical 3D cavity-transmon couplings, so that our results can be regarded as a step towards this hybrid cavity QED scheme.  

\bigskip
\textbf{Acknowledgements}:

We thank Silvia Ruffieux and Stefan Filipp for their contributions to the initial phase of the experiment and Bernhard Morath for manufacturing the cavity. 
We acknowledge the European Union H$2020$ FET Proactive project RySQ (grant N. $640378$). Additional support was provided by the Swiss National Science Foundation (SNSF) under project number $20020\_149216$ (FM) and by the National Centre of Competence in Research "Quantum Science and Technology" (NCCR QSIT), a research instrument of the SNSF.

\appendix

\section{Details on 3D cavity}
\label{app:cavity}
\begin{figure}[t] \centering \includegraphics[width=80mm]{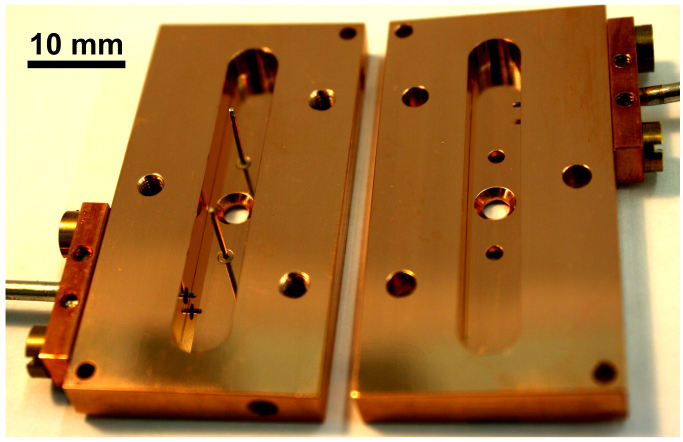} 	
\caption{Photograph of the rectangular copper 3D cavity with resonance frequency $\omega_{\text{c}} / 2\pi = 21.532~\text{GHz}$ ($\text{TE}_{301}$ mode). The atom entry and exit apertures as well as the intra cavity electrodes are clearly visible in the two halves of the cavity aligned by two alignment holes in the corners of each rectangle and closed by a set of screws. The microwaves are injected and extracted through two evanescently coupled microwave lines entering the cavity from the sides.  }
\label{fig:3Dcavity} 
\end{figure}

The rectangular cavity presented in Fig.~\ref{fig:3Dcavity} consists of two halves milled out from oxygen-free copper and its  dimensions ($w\times~h\times~l \simeq 42~\text{mm}\times6~\text{mm}\times8~\text{mm}$, rounded with radius $R=3~\text{mm}$ in the corners) determine the resonance frequency $\omega_{\text{c}} / 2\pi = 21.532~\text{GHz}$ of the $\text{TE}_{301}$ mode. Two $3$-mm-diameter holes allow the Rydberg atoms to enter and leave the cavity. One cylindrical electrode is mounted on each side of the atom beam at the nodes of the $\text{TE}_{301}$ mode to minimize perturbations of the mode structure and microwave losses. 
The transmission measurement uses two microwave antennas, that were adjusted to obtain a critically coupled $\text{TE}_{301}$ mode. In this configuration, the mode has a decay rate of $\kappa / 2\pi = 4.1~\text{MHz}$.

\section{Measurement noise and averaging}
\label{app:noiseave}

To measure the transmission of the cavity, we apply a weak probe tone to the cavity. The transmitted signal is amplified with a cryogenic, ultra-low-noise HEMT amplifier (Caltech Cryo1126) mounted at $T=3~\text{K}$. Low-pass and high-pass filtering at room temperature reduces the total noise power before further low-noise amplification. After heterodyne down-conversion using a double balanced mixer energized by a local oscillator, we obtain the signal at an intermediate frequency of $25~\text{MHz}$. To avoid aliasing, the signal is then low-pass filtered at the Nyquist frequency of $50~\text{MHz}$ before it is digitized by an analog-to-digital converter with $10~\text{ns}$ sampling interval. Afterwards, the signal is processed by a field-programmable gate-array (Xilinx Virtex4), where digital homodyne down-conversion and digital filtering with a $100~\text{ns}$ boxcar filter lead to measured time traces of the complex signal $\mathcal{A}(t)$ and its reference  $\mathcal{A}_0(t)$. Considering the noise added by the amplifiers in the detection chain, we can write each of the time traces as the sum of a signal component $\mathcal{A}_{\mathrm{S}} = I_{\mathrm{S}} + i Q_{\mathrm{S}}$ and a noise component $\mathcal{A}_{\mathrm{N}} = I_{\mathrm{N}} + i Q_{\mathrm{N}}$.

The noise background of the transmission measurement is characterized by the number of effective noise photons $n_{\mathrm{noise}}$ referred to the cavity output. For any detection chain, this number is calculated as the total noise power spectral density $\text{PSD}$ divided by the total gain $G$ and the photon energy $\hbar \omega$: $n_{\text{noise}}=\text{PSD} /  (G \hbar \omega)$. From measurements of the total gain and power spectral density, we have obtained $n_{\text{noise}}\simeq 34$, which is limited by the noise from the cryogenic HEMT amplifier. The SNR is then calculated as the ratio of integrated signal photon number $n_{\mathrm{c}} \kappa_{\mathrm{out}} \tau$ and effective noise photon number $\mathrm{SNR} = n_{\mathrm{c}} \kappa_{\mathrm{out}} \tau / n_{\mathrm{noise}}$.

To improve the SNR, we repeat the measurement at $25~\text{Hz}$ rate (limited by the pulse repetition rate of the UV laser) and first average $\mathcal{A}(t)$ and $\mathcal{A}_0(t)$ at the FPGA.
Special attention is paid in the analysis to cancel the effects of slow phase drifts of approximately $0.3~\text{deg.} / \text{min}$ between the probe tone and the local oscillator (probably induced by thermal drifts of the interferometer arms), which could lead to systematic errors of the extracted amplitudes and phases. We minimize these errors by averaging only $100$ cycles at the FPGA (\textit{i.e.}, $4~\text{s}$ of integration). We thus obtain the averaged complex amplitudes:   
\begin{align}
  \overline{\mathcal{A}} &=  A_{\mathrm{S}} e^{i \; \phi_{\mathrm{S}}} + \overline{A_{\mathrm{N}}} e^{i \; \overline{\phi_{\mathrm{N}}}} \\
\overline{\mathcal{A}_{0}} &= A_{0,\mathrm{S}} e^{i \; \phi_{0,\mathrm{S}}} + \overline{A_{0,\mathrm{N}}} e^{i \; \overline{\phi_{0,\mathrm{N}}}} \ . \nonumber
\end{align}

To extract the phase change $\delta \phi=\phi_{\mathrm{S}}-\phi_{0,\mathrm{S}}$, we average the Hermitian inner product of the averaged complex amplitudes: 
\begin{align}
\langle \overline{\mathcal{A}}\cdot \overline{\mathcal{A}_0}^{\; *} \rangle &= A_{\mathrm{S}} A_{0,\mathrm{S}} e^{i \delta \phi} + A_{\mathrm{S}} e^{i \phi_{\mathrm{S}}} \langle \overline{\mathcal{A}_{0,\mathrm{N}}}^{\; *} \rangle  \\
&+ A_{0,\mathrm{S}} e^{-i \phi_{0,\mathrm{S}}} \langle \overline{\mathcal{A}_{\mathrm{N}}} \rangle + \langle \overline{\mathcal{A}_{\mathrm{N}}} \rangle \langle \overline{\mathcal{A}_{0,\mathrm{N}}}^{\; *} \rangle \ .  \nonumber
\end{align}
The averaged noise terms $ \langle \overline{\mathcal{A}_{\mathrm{N}}} \rangle$ and $\langle \overline{\mathcal{A}_{0,\mathrm{N}}}^{\; *} \rangle$ decay to zero for a large number of averages and we extract the phase change from the only remaining term.

To obtain the amplitude change, we calculate the squared absolute values of the complex amplitudes 
\begin{align}
\langle {\lvert\overline{\mathcal{A}}\rvert}^2 \rangle &= {A}_{\mathrm{S}}^2  + \langle \overline{{A}_{\mathrm{N}}}^2 \rangle + 2 \langle I_{\mathrm{S}} \overline{I_{\mathrm{N}}} \rangle + 2 \langle Q_{\mathrm{S}} \overline{Q_{\mathrm{N}}} \rangle  \\
\langle {\lvert\overline{\mathcal{A}_0}\rvert}^2 \rangle &= {A}_{0,\mathrm{S}}^2  +  \langle \overline{{A}_{0,\mathrm{N}}}^2 \rangle + 2 \langle I_{0,\mathrm{S}} \overline{I_{0,\mathrm{N}}} \rangle + 2 \langle Q_{0,\mathrm{S}} \overline{Q_{0,\mathrm{N}}} \rangle \  .  \nonumber
\end{align} 
Because the signal and noise components are uncorrelated, the third and fourth terms are proportional to $\langle \overline{I_{\mathrm{N}}} \rangle$ and $\langle \overline{Q_{\mathrm{N}}} \rangle$, respectively, and vanish for sufficient averaging. In both expressions the second term  converges to the same nonzero value related to the total noise power $P=\overline{{A}_{\mathrm{N}}}^2 / 2Z$, where $Z=50~\Omega$ is the line impedance. 
This value is determined separately from a measurement with switched-off probe tone and allows us to extract the signal amplitudes ${A}_{\mathrm{S}}$ and ${A}_{\mathrm{0,S}}$.
We divide the difference between signal and reference amplitude by the signal amplitude $A_{0,\mathrm{S}}(\omega_{\mathrm{c}})$ at the cavity resonance to obtain the change in normalized amplitude: 
\begin{equation}
\delta \!A_{\text{n}}=\frac{{A}_{\mathrm{S}}-{A}_{\mathrm{0,S}}}{A_{0,\mathrm{S}}(\omega_{\mathrm{c}})} \  .
\end{equation}

The changes of the amplitudes and phases were determined as described above by averaging over $5 \cdot 10^4$ cycles, \textit{i.e.}, over $100$ cycles at the FPGA followed by averaging over $500$ FPGA outputs. We have noticed a small artificial phase offset of $-0.104(4)$ degree between signal and reference traces (measured in the absence of atoms) that we subtract from the measured phase change.

\section{Full temporal and spectral data compared to fits}
\label{app:fulldata}

\begin{figure}[t] \centering \includegraphics[width=85mm]{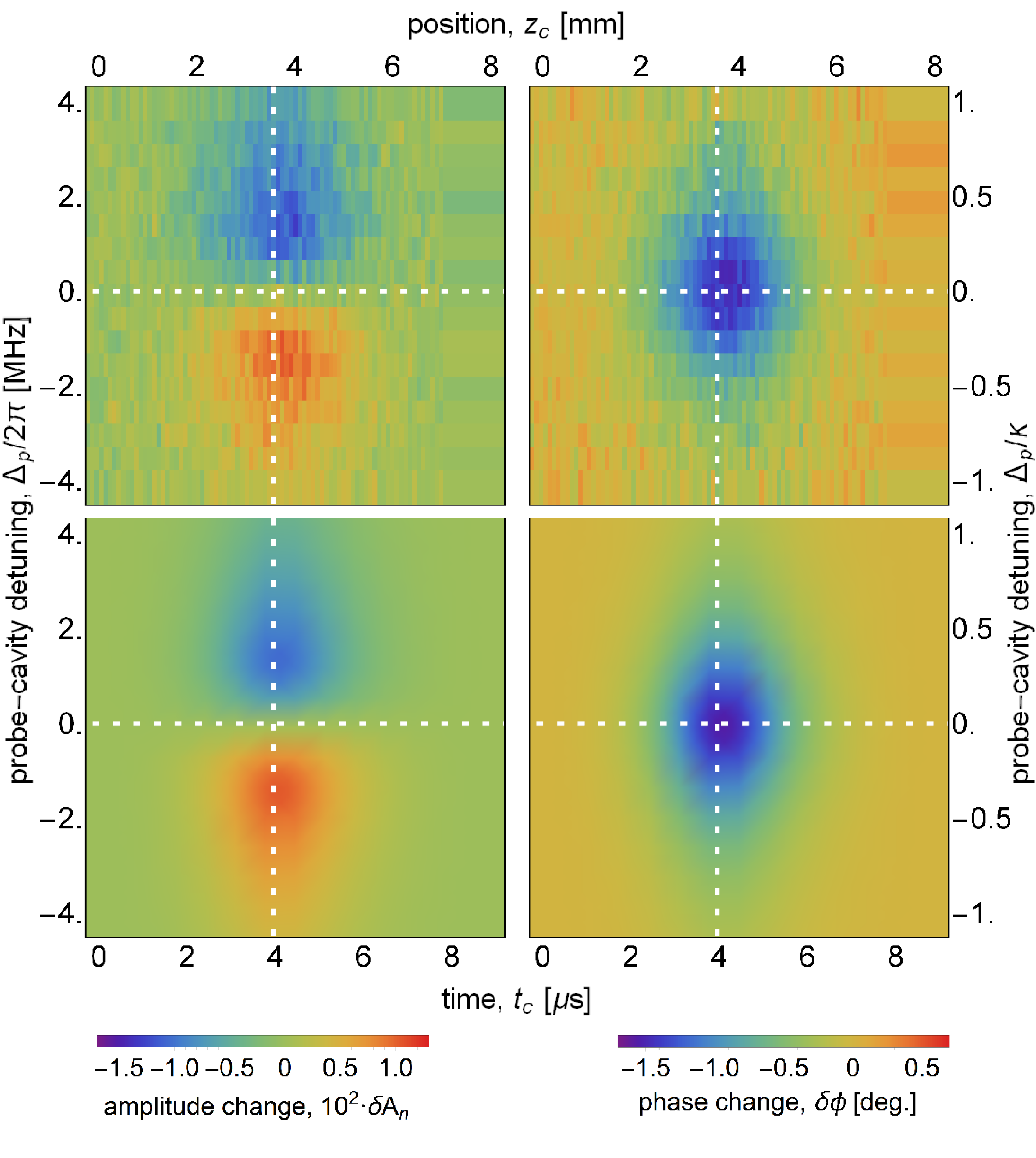} 	
\caption{The top panels show the measurements of change in amplitude $\delta \! A_{\text{n}}$ (left) and phase $\delta \phi$ (right) as functions of the position along the cavity axis $z_{\mathrm{c}}$ (and corresponding time $t_{\text{c}}$)  and probe-cavity detuning $\Delta_{\text{p}}$. The bottom panels show the corresponding fits of the change in amplitude (left) and phase (right) with the complete model introduced in the main text. The white dashed lines indicate the cuts through the data and fits at $\Delta_{\text{p}} / 2\pi=0~\text{MHz}$ and $t_{\text{c}}=t_{\text{min}}$ that are shown in Fig.~\ref{fig:modelplot}(a,b) and Fig.~\ref{fig:freqscanplot}, respectively.  }
 \label{fig:2Ddata}
 \end{figure}

 In Fig.~\ref{fig:2Ddata} we present the full data set (top panels) and the fit (bottom panels) of the changes in amplitude $\delta \! A_{\text{n}}$ (left) and phase $\delta \phi$ (right) as functions of the position along the cavity axis $z_{\mathrm{c}}$ (and corresponding time $t_{\text{c}}$) and the probe-cavity detuning $\Delta_{\text{p}}$. 
The cuts through the data and fits at zero probe-cavity detuning and around the position of minimal atom-cavity detuning were used in Section \ref{ssec:results2} to discuss the time dependence of the dispersive shift and the effect of the dispersive shift on the transmission spectrum, respectively. 
We fit the full data set with a model that takes into account the calculated single-atom coupling $g_{1}(z_{\text{c}})$ and the measured atom-cavity detuning $\Delta_{\text{a}}(t_{\text{c}})$ (see Fig.~\ref{fig:modelplot}(c) and (d)). The fit determines the origin of the time scale ($t_{\text{c}}=0~\mu s$) that is used to align $\delta \! A_{\text{n}}(t_{\text{c}})$ and $\delta \phi(t_{\text{c}})$ with the cavity frame and the number of coupled Rydberg atoms $N =3.3(2) \cdot 10^3 $ that corresponds to a maximal collective coupling of $g_{N\text{,max}} / 2\pi=1.01(2) ~\text{MHz}$. The quantitative agreement between fit and data implies the observation of a collective dispersive shift, the time dependence of which is determined by the time-dependent collective coupling $g_{N}(t_{\text{c}})$ and atom-cavity detuning $\Delta_{\text{a}}(t_{\text{c}})$.

\section{Critical photon number}
\label{app:ncrit}

\begin{figure}[b] \centering \includegraphics[width=80mm]{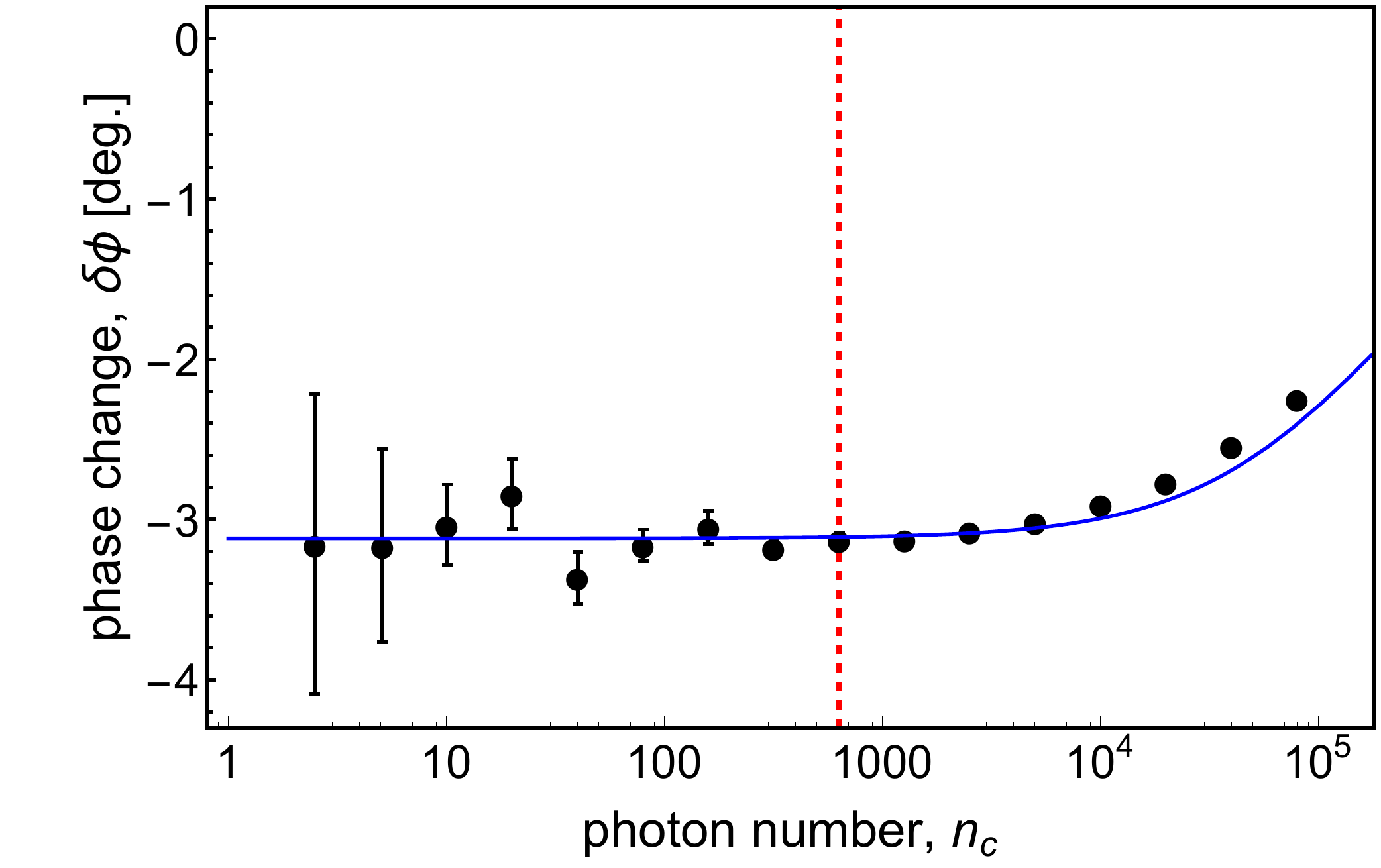} 	
	\caption{Measured phase change $\delta \phi$ (black points) at the cavity resonance as a function of the cavity photon number $n_{\mathrm{c}}$. The blue line is calculated according to Eq.~\ref{eq:chiNfuntionnc}. The dashed red line indicates the cavity photon number that was used for the measurements presented in Section~\ref{sec:results} and Appendix~\ref{app:fulldata}.}
	\label{fig:power}
\end{figure}

 The eigenvalues of the Jaynes-Cummings Hamiltonian~\cite{Jaynes1963} for a single qubit coupled to a resonator are $E_{\pm,n_{\mathrm{c}}} / \hbar = n_{\mathrm{c}} \omega_{\mathrm{c}} \pm  0.5 \sqrt{\Delta_{\mathrm{a}}^2 + 4 g_1^2 n_{\mathrm{c}} }$ for a given photon number $n_{\mathrm{c}}$. In the dispersive limit, the dressed cavity transition frequency is given by the difference in eigenenergy of states that differ by one photon and have the same qubit state (ground and excited states correspond to $-$ and $+$, respectively). For a qubit in the ground state and in the limit $n_{\mathrm{c}}\gg 1$, this leads to:
\begin{align}
\chi_1 \left(n_{\mathrm{c}} \right) &\simeq \frac{\mathrm{d}}{\mathrm{d} n_{\mathrm{c}}}\left(\frac{-\sqrt{\Delta_{\mathrm{a}}^2 + 4 g_1^2 n_{\mathrm{c}}}}{2}\right) \nonumber \\
 &\simeq -\frac{g_1^2}{\Delta_{\mathrm{a}}} \frac{1}{ \sqrt{ 1 + n_{\mathrm{c}} / n_{\mathrm{crit}}}} \ .
\label{eq:chi1funtionnc}
\end{align} 
In Eq.~\ref{eq:chi1funtionnc}, the critical photon number $n_{\mathrm{crit}} = \Delta_{\mathrm{a}}^2 / 4 g_1^2$ gives a scale for which (in the two-level approximation) the dispersive limit breaks down~\cite{Blais2004}.
One can characterize the dependence of the dispersive shift on the photon number more carefully, especially taking into account the dephasing that occurs for large photon numbers~\cite{Boissonneault2008}; $n_{\mathrm{crit}}$ nevertheless provides the characteristic scale. 

Exact diagonalization of the Tavis-Cummings Hamiltonian is more complicated~\cite{Tavis1968}. In the limit where the number $n$ of excitations (number of photons and of atoms in the excited qubit state) in the system is large compared to the number of atoms $N$, the eigenenergies can be approximated~\cite{Narducci1973,Garraway2011} by $E_{j,n} / \hbar \simeq n \omega_{\mathrm{c}} + j \sqrt{\Delta_{\mathrm{a}}^2 + 4 g_1^2 n}$. Here, $-N/2 \leq j \leq N / 2$ corresponds, in the dispersive limit, to the polarization of the atomic ensemble $j= \langle J_z \rangle$. Thus, we can extract the dependence of the cavity dispersive shift on the photon number:
\begin{align}
\chi_N \left(n_{\mathrm{c}} \right) &\simeq \frac{\mathrm{d}}{\mathrm{d} n} 
\left(j \sqrt{\Delta_{\mathrm{a}}^2 + 4 g_1^2 n } \right) \nonumber \\
&\simeq \frac{g_1^2}{\Delta_{\mathrm{a}}} 2 \langle J_z \rangle\frac{1}{ \sqrt{ 1 + n_{\mathrm{c}} / n_{\mathrm{crit}}}} \ ,
\label{eq:chiNfuntionnc}
\end{align} 
where we have used $n = n_{\mathrm{c}} + j + N / 2 \simeq n_{\mathrm{c}}$, which is valid for $n \gg N$. Thus, $n_{\mathrm{crit}}$ still provides the threshold around which the dispersive shift starts to decrease.

In our experiments, the number of the probe microwave photons should have a negligible effect on the dispersive shift, because $n_{\mathrm{c}} \simeq 600 \ll n_{\mathrm{crit}} \simeq 10^5$. 
To verify this, we have measured the phase change for a resonant probe as a function of the number of photons in the cavity. As shown in Fig.~\ref{fig:power}, a significant decrease in dispersive shift is only observed at cavity photon numbers that are at least one order of magnitude higher than the photon numbers used to obtain the results presented in Section \ref{sec:results}. The data agree qualitatively with the scaling predicted with Eq.~(\ref{eq:chiNfuntionnc}) and the previously calculated value of  $n_{\mathrm{crit}}$. We attribute the deviation at large photon numbers to the uncertainty in the photon number and to a reduction of the dispersive shift induced by the ac Stark shift.

\section{Optimal atom detection}
\label{app:SNR}

In the limit of large power SNR, the uncertainty in the phase $\phi$ resulting from the Gaussian noise for a single-shot transmission measurement on resonance is given by $\sigma_\phi = 1 / \sqrt{SNR}$ (by taking the ratio of noise and signal amplitudes and assuming $\tan \sigma_\phi \simeq \sigma_\phi$ for $\mathrm{SNR} \gg 1$). The uncertainty of the phase change $\delta\phi = \phi - \phi_0$ is then $\sigma_{\delta\phi} =  1 / \sqrt{SNR} \sqrt{2+(2 \chi / \kappa)^2}$, which takes into account the reduction of the transmitted power for the shifted resonance. This expression matches the uncertainty of the phase measured in our experiment within $ 20\% $ after multiplying by $1 / \sqrt{k}$ to take into account the $k$ averages. 

The atom number is extracted as $N =  \Delta_{\mathrm{a}} \kappa / 2 g_1^2 \cdot \tan(\delta\phi)  $ and thus has a relative uncertainty $\sigma_N / N = (\tan(\delta\phi) + 1/\tan(\delta\phi)) \sigma_{ \delta\phi} $. After inserting the phase change $\delta\phi = \arctan (2 \chi / \kappa)$ and its single-shot uncertainty, we obtain: 
\vspace{-1mm}
\begin{equation}
\frac{\sigma_ N}{N} = \left(\frac{2 \chi}{\kappa} + \frac{\kappa}{2 \chi}\right) \sqrt{2+\left(\frac{2 \chi}{\kappa}\right)^2} \frac{1}{\sqrt{SNR}} \ .
\end{equation}
This function clearly depends on the ratio between the dispersive shift and the resonator linewidth $x=2\chi / \kappa$ and reaches its minimum $\sigma_N / N \simeq 3.33/\sqrt{\mathrm{SNR} }$ for $x_{\mathrm{opt}} \simeq 0.8 $.

\end{document}